\documentclass[aps,prl,twocolumn,superscriptaddress,showkeys]{revtex4-2}

\usepackage[english]{babel}

\UseRawInputEncoding

\usepackage{amsmath}
\usepackage{graphicx}
\usepackage{xr-hyper}
\usepackage[colorlinks=true, allcolors=blue]{hyperref}
\usepackage[many]{tcolorbox}

\usepackage{romannum}

\graphicspath{{main_figures/}}

\def \suppsec[#1]{Supplementary Material Section}
\def \SF[#1]{Suppl.~Fig.~#1}
\def \F[#1]{Fig.~#1}

\def \todo[#1]{\color{red}#1\color{black}}

\newtcolorbox{boxI}{
    colback = sub, 
    colframe = main, 
    boxrule = 0pt, 
    toprule = 6pt 
}

\renewcommand{\v}[1]{\ensuremath{\mathbf{#1}}}

\makeatletter
\newcommand*{\addFileDependency}[1]{
  \typeout{(#1)}
  \@addtofilelist{#1}
  \IfFileExists{#1}{}{\typeout{No file #1.}}
}
\makeatother

\newcommand*{\myexternaldocument}[1]{%
    \externaldocument{#1}%
    \addFileDependency{#1.tex}%
    \addFileDependency{#1.aux}%
}

\myexternaldocument{suppl}

\begin{document}
\title{Delusive chirality and periodic strain pattern in moir\'{e} systems }

\author{\'{A}rp\'{a}d P\'{a}sztor}
\email{arpad.pasztor@unige.ch}
\affiliation{DQMP, Universit\'e de Gen\`eve, 24 Quai Ernest Ansermet, CH-1211 Geneva, Switzerland}
\affiliation{These authors contributed equally to this work.}

\author{Ishita Pushkarna}
\affiliation{DQMP, Universit\'e de Gen\`eve, 24 Quai Ernest Ansermet, CH-1211 Geneva, Switzerland}
\affiliation{These authors contributed equally to this work.}

\author{Christoph Renner}
\email{christoph.renner@unige.ch}
\affiliation{DQMP, Universit\'e de Gen\`eve, 24 Quai Ernest Ansermet, CH-1211 Geneva, Switzerland}

\begin{abstract}
Geometric phase analysis (GPA) is a widely used technique for extracting displacement and strain fields from scanning probe images. Here, we demonstrate that GPA should be implemented with caution when several fundamental lattices contribute to the image, in particular in twisted heterostructures featuring moir\'e patterns. We find that in this case, GPA is likely to suggest the presence of chiral displacement and periodic strain fields, even if the structure is completely relaxed and without distortions. These delusive fields are subject to change with varying twist angles, which could mislead the interpretation of twist angle dependent properties. 
\end{abstract}


\maketitle

\section{Introduction}
Geometric phase analysis (GPA), the analysis of the spatial variation of the phase of a periodic signal, like in a high-resolution image obtained with a scanning tunneling microscope (STM), has been widely applied to gain insight into the physics of correlated electron systems. Specific examples include phase-sensitive identification of a \textit{d}-form factor density wave in cuprates \cite{Fujita2014}, the observation of a multiband character of the charge density wave (CDW) in a transition metal dichalcogenide \cite{Pasztor2021}, and discommensuration and topological defects in ordered electronic phases \cite{Mesaros2011,Mesaros2016,Okamoto2015,Pasztor2019}. GPA is also extensively deployed to correct image distortions \cite{Lawler2010,Zeljkovic2014}, to extract chiral CDW displacement fields \cite{Singh2022}, to quantify local moir\'e lattice heterogeneity \cite{Benschop2021}, spatial variation of strain \cite{Zeljkovic2015,Gao2018,Walkup2018}, and periodic strain modulations in van der Waals heterostructures \cite{Zhao2022}.

\begin{figure}[bb]
    \centering
    \includegraphics[width=\columnwidth]{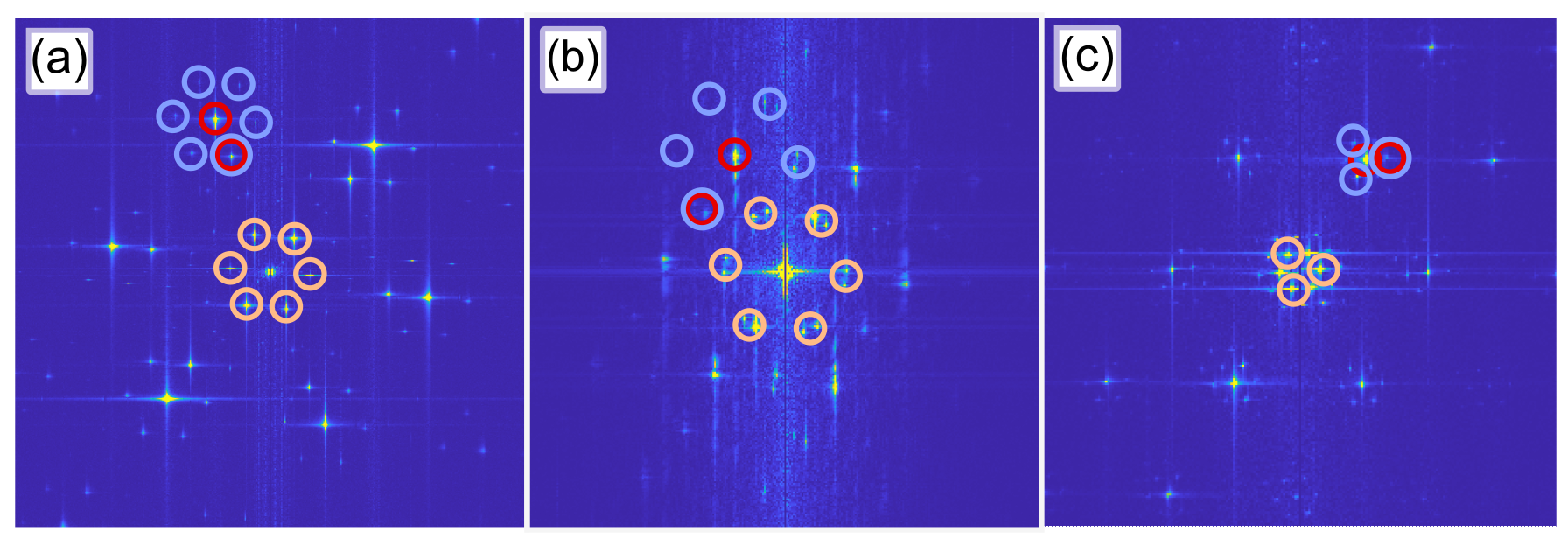}
    \caption{Fourier transforms of STM images of different heterostructures with moir\'e patterns. (a) Graphene on WSe$_2$ (1\textdegree~twist angle) \cite{Sun2022, Sun2022_2}, (b) monolayer MoS$_2$ on 2H-NbSe$_2$ (30\textdegree, at 77~K($>T_{CDW}$) (adapted with permission from \cite{Martinez2018}, copyright 2018 American Chemical Society), and (c) monolayer MoS$_2$ on Au(111) (7.7\textdegree) \cite{Pushkarna2023}. Red, blue, and orange circles highlight lattice, satellite, and moir\'e peaks, respectively. } 
    \label{fig:FT_heterostructures}
\end{figure}

GPA has been implemented in different ways: using spatial lock-in (SL), also known as \textit{coarse-grained phase field extraction} or \textit{Lawler-Fujita method} \cite{Lawler2010,Benschop2021,Savitzky2017,Baggari2018} (see also \suppsec[]~\Romannum{3}), using real-space fitting \cite{Pasztor2019}, and combining Fourier-filtering with intensity thresholding \cite{Singh2022}. These methods all extract the spatial variation of the phase with respect to a reference signal corresponding to the center of an atomic or a CDW Bragg peak in Fourier space. They all involve a filtering step performed either in Fourier space, by masking a region of interest and inverse transforming, or in real space, by convoluting the raw data with a smoothing function. These filterings are equivalent and have a decisive impact on the phase field obtained. 

Here, we discuss the limitations and potential pitfalls of GPA, focusing on deceptive deformation fields and non-existent chiral distortions that can appear. We show that ambiguity arises when the analyzed image is formed by several non-overlapping fundamental lattices. This is particularly relevant to STM images of moir\'e patterns.

\section{Experiments and Model}
Moir\'e patterns are commonly observed by STM when imaging two periodic structures with different lattice orientations and/or different lattice constants \cite{Benschop2021, Martinez2018,Zhang2017,Zhang2021NatComm,Pan2018,Jiang2019,Shabani2021,Zhao2022,Sun2022}. In Figure~\ref{fig:FT_heterostructures}, we show the Fourier transforms (FTs) of three high-resolution STM images of moir\'e patterns obtained on different heterostructures: graphene on WSe$_2$ (Figure~\ref{fig:FT_heterostructures}(a)), monolayer MoS$_2$ on 2H-NbSe$_2$ (Figure~\ref{fig:FT_heterostructures}(b)) and monolayer MoS$_2$ on Au(111) (Figure~\ref{fig:FT_heterostructures}(c)). A common feature of all three FTs is a high degree of wavevector mixing, with many intense peaks besides the fundamental lattice peaks ($\v{q}_{Li}$). These additional $\v{q}$-vectors correspond to a linear combination of the constituent lattices wavevectors: $\v{q}=\sum_i n_i\v{q}_{Li}$, where $n_i$ are integers. Ultimately, we can generate all the FT components by choosing specific sets of $\{n_i\}$ coefficients. The peaks corresponding to the lowest order wavevector mixing, i.e. with the most zero components in $\{n_i\}$, will typically appear brightest. In Figure~\ref{fig:FT_heterostructures}, example lattice peaks corresponding to $\v{q}_{Li}$ are marked by red circles, while the shortest moir\'e wavevectors $\v{q}_{Mj}$, which correspond to the differences between the nearest $\v{q}_{Li}$ of the two lattices generating the moir\'e pattern, are outlined by orange circles. Blue circles highlight example \textit{satellite} or moir\'e \textit{replica} peaks, whose wavevectors correspond to first-order linear combinations of $\v{q}_M$ and $\v{q}_L$ components. Since $\v{q}_{Mj}$ are the shortest wavevectors in the system (in the absence of other periodicities than the constituent lattices), these satellite peaks are the closest to the lattice peaks. Note that one of the satellite peaks always coincides with one of the lattice peaks and that the lattice peaks corresponding to the top layer in a heterostructure are typically more intense. For clarity, we only highlight some of the peaks in Figure~\ref{fig:FT_heterostructures}. 

To investigate possible artifacts of the GPA technique, we design numerically generated real-space images whose FT intensity maps contain all the features discussed above. The simplest model consists in a sum of plane waves with properly defined wavevectors. To simulate the three-fold symmetry of the heterostructures shown in Figure~\ref{fig:FT_heterostructures}, we define the $\v{q}$-vectors by considering three non-equivalvent ($2\pi/3$ rotated) lattice peaks $\v{q}_{Li}$ ($i=1,2,3$) and their satellites $\v{q}_{ij}=\v{q}_{Li}+\v{q}_{Mj}$ ($j=1,2...N$). For a uniform notation, we define the lattice peaks as $\v{q}_{i0}=\v{q}_{Li}$, indeed $\v{q}_{M0}=0$. In the following simulations, we allow up to six satellite peaks ($N=6$) placed symmetrically around each lattice peak to best mimic our data. Note that in general, any number of (satellite) peaks could be included. However, we are only interested in those that are the closest to the lattice peaks as all the others are filtered out when applying GPA to real data. It is for the same reason that we do not consider the peaks ($\v{q}_{Mj}$) close to the origin (\v{q}=0).  Once the wavevectors of the first lattice peak and of its six satellites have been defined, e.g. $\v{q}_{1j}$, we obtain the others by $2\pi/3$ rotations around the origin to account for the spatial symmetry of the systems we are simulating. The corresponding topographic signal can then be calculated as:
\begin{equation}
\begin{split}
T(\v{r})&=\sum_{i=1}^{3}T_i(\v{r})=\sum_{i=1}^{3}\sum_{j=0}^NA_{ij}(\v{r})\sin(\v{q}_{ij}\v{r}+\vartheta_{ij}),
\end{split}
\label{eq:beating2}
\end{equation}
where $T_i(\v{r})$ is the signal from the $i$-th lattice peak and its satellites. To simplify the discussion we set $\vartheta_{ij}=0$ without impacting our final conclusions. It allows a convenient description in terms of peak intensities in the FT (which are determined by the amplitudes $A_{ij}$ of the plane waves in real space). Considering $\vartheta_{ij}\neq0$ can give a more general description with additional complexity that may be worth investigating in specific systems. Using the identities given in \suppsec[]~\Romannum{8}, we obtain
\begin{equation}
\begin{split}
T_i(\v{r})=\sum_{j=0}^N A_{ij}(\v{r})\sin(\v{q}_{ij}\v{r})=\\
=\sum_{j=0}^N A_{ij}(\v{r})\sin(\v{q}_{Li}\v{r}+\v{q}_{Mj}\v{r})=
A_{i}(\v{r})\sin(\v{q}_{Li}\v{r}+\varphi_i(\v{r})).
\end{split}
\label{eq:beating2}
\end{equation}
The terms $A_i(\v{r})$ and $\varphi_i(\v{r})$ satisfy: 
\begin{equation}
\begin{split}
A_i^2(\v{r})&=\sum_{k,l}A_{il}(\v{r})A_{ik}(\v{r})\cos(\Delta\v{q}_{Mkl}\v{r})
\end{split}
\label{eq:beating2_ampl_phase}
\end{equation}
and
\begin{equation}
    \tan(\varphi_i(\v{r}))=\frac{\sum_jA_{ij}(\v{r})\sin(\v{q}_{Mj}\v{r})}{\sum_jA_{ij}(\v{r})\cos(\v{q}_{Mj}\v{r})}
    \label{eq:phase_sum_many_sine}
\end{equation}
where $\Delta\v{q}_{Mkl}\equiv \v{q}_{Mk}-\v{q}_{Ml}$.

This is an amplitude-modulated signal, which is a spatial analog of the beating phenomenon well-known from acoustics. In real space images, it appears as a periodic domain structure defined by the $\v{q}_{Mj}$ wavevectors, similar to a moir\'e pattern (\suppsec[]~\Romannum{1}).
\begin{figure}[h!]
    \centering
    \includegraphics[width=\columnwidth]{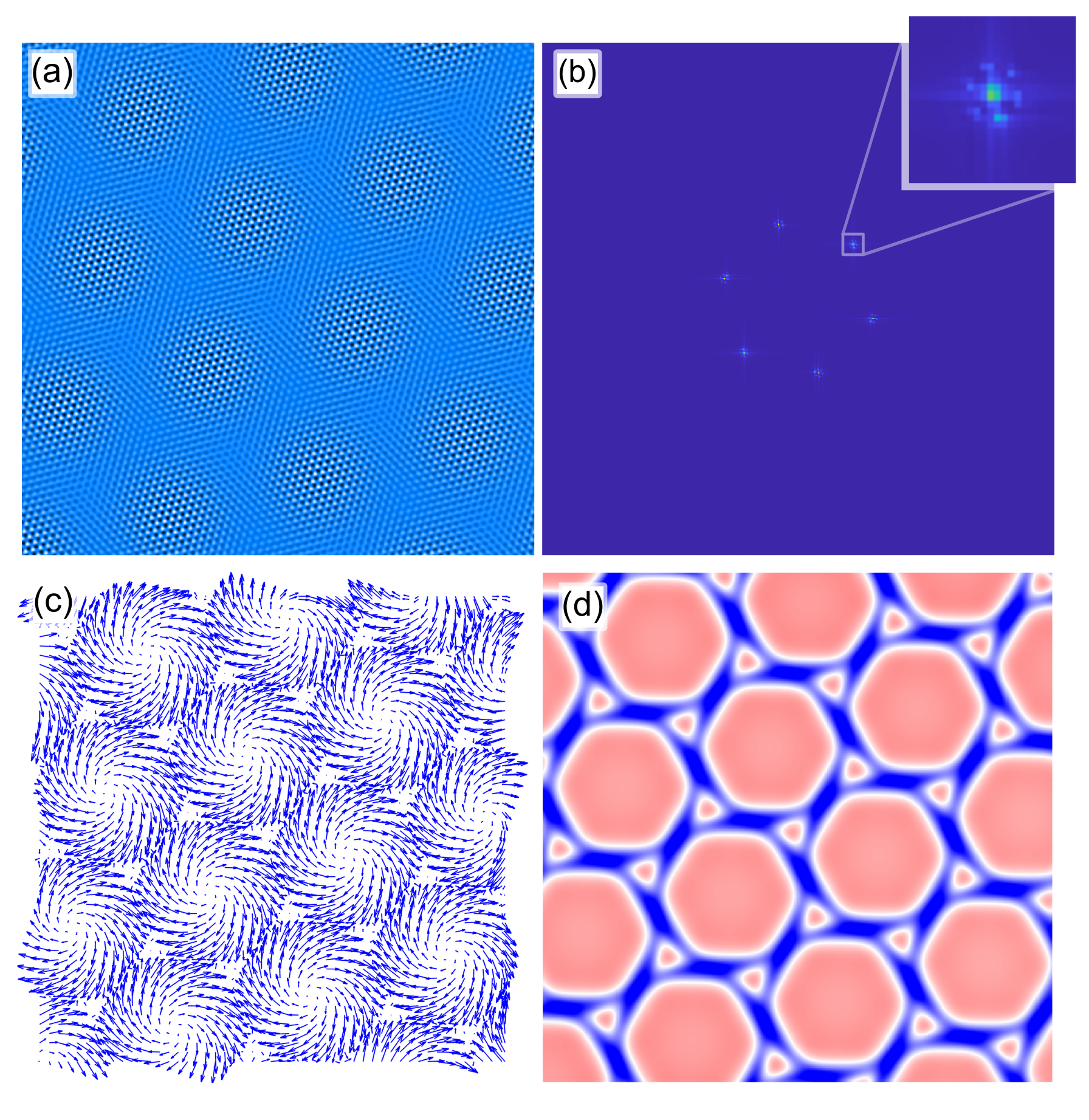}
    \caption{Chiral displacement field and periodic strain in the sum of plane waves model. (a) Calculated topography showing a domain structure due to spatial beating. (b) FT of (a). Inset is a magnification of one lattice peak with its six satellites (note that one of them is brighter). (c) Displacement vector field and (d) periodic biaxial strain field obtained from the phase fields of three main lattice peaks and their satellites by evaluating Equation~\ref{eq:phase_sum_many_sine}.}
    \label{fig:sum_of_sines_chirality}
\end{figure}

Equation (\ref{eq:beating2}) describes the aggregated signal of central and surrounding satellite peaks, and its spatially varying phase (Equation~\ref{eq:phase_sum_many_sine}). We can explicitly calculate this phase field and the corresponding displacement field, which then allows us to extract a strain field as described in \suppsec[#1]~\Romannum{2}. In the next section, we do this in two ways. First, we use the phase field obtained from the three main lattice peaks and their satellites to compute the displacement field, solving an over-determined system of equations in the least square sense (\suppsec[#1]~\Romannum{2}). Second, we show what happens if we only use pairs of two lattice peaks and their satellites, which is in principle sufficient to describe any real distortion in the two-dimensional STM image plane. 

\begin{figure*}[ht]
    \centering
    \includegraphics[width=6in]{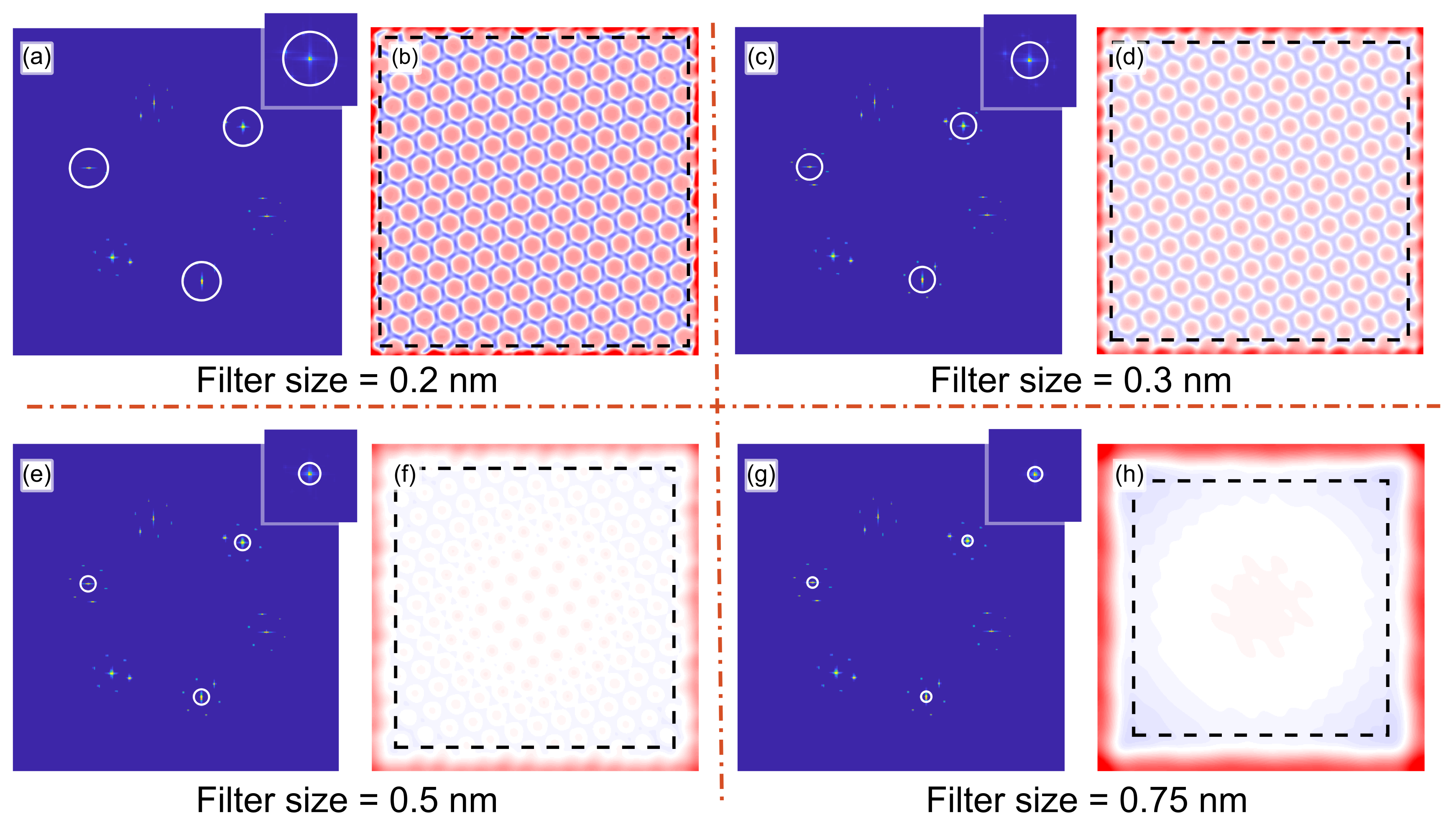}
    \caption{Periodic biaxial strain field extracted from a model moir\'e structure using the SL method as a function of decreasing filter size. (a), (c), (e) and (g) show the same FT of a sum of plane waves topographic image. The filter size used to extract the strain field is depicted as a white circle (with a diameter of $\sigma$ of the Gaussian filter in Fourier space), and its effect on the satellite peak intensity is shown in each inset. (b), (d), (f) and (h) show the corresponding periodic biaxial strain field in real space extracted using the SL method for each given filter size. The dashed square delimits the region $3\sigma$ (in real space) away from the edges of the strain maps, where the filtering kernel is entirely within the image and edge effects disappear. Note the vanishing of the strain field when suppressing the satellite peaks with a small filter. The large filters yield the same result as the one obtained analytically in \SF[7].}
    \label{fig:tightening_filter_sum_of_sines}
\end{figure*}

\section{Discussion}

The key elements influencing the GPA are the satellite peaks,  which ultimately determine the nature of possible displacement and strain fields. When all satellite peaks have the same intensity, the corresponding topography shows domains due to the amplitude beating (Equation~\ref{eq:beating2_ampl_phase}), but the displacement field is strictly zero (Equation~\ref{eq:phase_sum_many_sine}, \SF[3]). However, for any asymmetry in the satellite peak intensities, we not only find an amplitude beating that shapes a domain structure in real space, but also periodically varying displacement and strain fields as shown in \F[\ref{fig:sum_of_sines_chirality}] and in Suppl. Figs. 4 and 5. These fields are modulated with a periodicity of $2\pi/q_{M}$ and their fine structure depends on the details in the satellite peak intensities as discussed below.

To give an insight into the variety of possible displacement and strain fields depending on the satellite peak structure, we consider a simple case where all the satellite peaks have the same intensity, except for one (the $k$-th) which is stronger: $A_{ij}=C<A_{ik}$ $\forall i$ and $\forall j\neq k$ ($j,k\neq 0$).  When the wavevector corresponding to the stronger satellite peak ($\v{q}_{ik}$)  is parallel with $\v{q}_{Li}$, Equation~\ref{eq:phase_sum_many_sine} yields a radial displacement field. This field is pointing away from (towards) the center of the beating domain when the satellite peak wavevector is shorter (longer) than $\v{q}_{Li}$, corresponding to tensile (compressive) strain within each domain (see \SF[4]). 

Whenever $\v{q}_{ik}$ is not parallel with $\v{q}_{Li}$, Equation~\ref{eq:phase_sum_many_sine} yields a chiral displacement field whirling around the center of each domain (\F[\ref{fig:sum_of_sines_chirality}]). The left-handed or right-handed vorticity depends on the angle between $\v{q}_{ik}$ and $\v{q}_{Li}$ (see \F[\ref{fig:sum_of_sines_chirality}]  and \SF[5]). Note that this chiral displacement field is zero in the middle of the domain and increases toward the domain edges, exactly as observed for the chiral CDW displacements in TaS$_2$~\cite{Singh2022}.

The analysis so far was purely analytical, evaluating Equation~\ref{eq:phase_sum_many_sine} for a given set of $A_{ij}$ amplitudes and $\v{q}_{ij}$ wavevectors. It shows that whenever the satellite peak structure has an inhomogeneous structure, delusive displacement, and strain fields appear in a system that is perfectly relaxed and undistorted by construction. Such inhomogeneities can be present for several reasons in real STM images of moir\'e structures, but in any case, because all satellite peaks are not equivalent since one of them always coincides with a lattice peak. Consequently, there is a significant risk of misinterpreting experimental data in terms of displacement fields and strain which are not present in the actual material.

To illustrate possible strain misinterpretations further, we perform GPA by applying the popular SL technique used to restore and analyse experimental STM data to our numerically generated images. This approach enables us a perfect control over the input parameters and assess their impact on the GPA. We indeed observe the same effects, finding finite periodic displacement and strain fields (\SF[6]) although the input structure does not have any of them. Unsurprisingly, since the displacement and strain fields depend on the satellite peaks, the response depends on the filter size applied to the data for the SL procedure. \F[\ref{fig:tightening_filter_sum_of_sines}] shows how the displacement field progressively disappears when reducing the Fourier space included in the filter around the lattice peaks at $\v{q}_{Li}$. Likewise, the displacement field will progressively vanish if the satellite peaks shift further away from the lattice peak while using a constant filter size. The moir\'e satellite peaks move away from the lattice peaks in heterostructures with increasing twist angle, and the resulting vanishing displacement field can be mistaken for a twist angle dependent reduction of strain \cite{Zhao2022}. 

\begin{figure}[htp]
    \centering
    \includegraphics[width=\columnwidth]{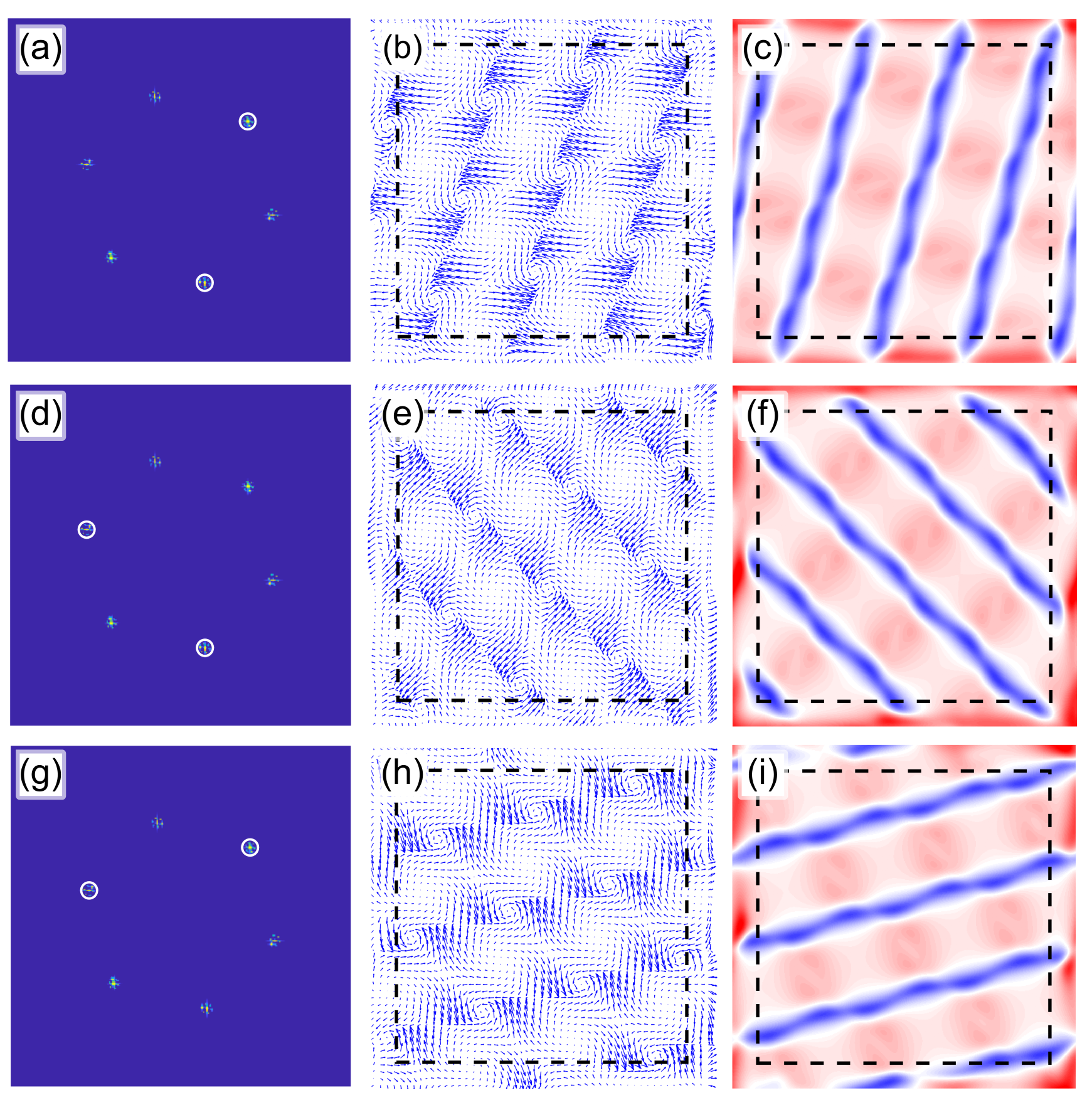}
    \caption{Displacement and strain field artifacts emerging in calculated images which include wavevector mixing. (a), (d) and (g) show the same FT of the calculated topography, with highlighted different pairs of lattice peaks and their satellite peaks (enclosed in the filter) used to extract the corresponding displacement fields shown in panels (b), (e) and (h), and strain fields shown in panels (c), (f) and (i). The white circles indicate the filter size used for the GPA (with a diameter of $\sigma$ of the Gaussian filter in Fourier space), and the dashed squares delimit the region $3\sigma$ (in real space) away from the edges of the images, where the filtering kernel is entirely within the image and edge effects disappear. Note the remarkable dependence of the results on the selected pairs of lattice peaks and the appearance of a spurious stripe texture. }
    \label{fig:sum_of_sines_2q}
\end{figure}

An additional analytical element we consider in the following is that two linearly independent in-plane vectors should uniquely define any two-dimensional displacement. Hence, we should find the same displacement field for any choice of lattice peaks pair and their satellites (see \suppsec[]~\Romannum{2}). However, comparing (either the SL or the analytical) results obtained with the three possible pairs of $\v{q}$-vectors for our undistorted model heterostructure, we observe three very different patterns with symmetry-lowering strain and displacement fields in \F[\ref{fig:sum_of_sines_2q}] (and in \SF[8]{}), providing additional evidence that they are not real.  

\begin{figure}[tp]
    \centering
    \includegraphics[width=\columnwidth]{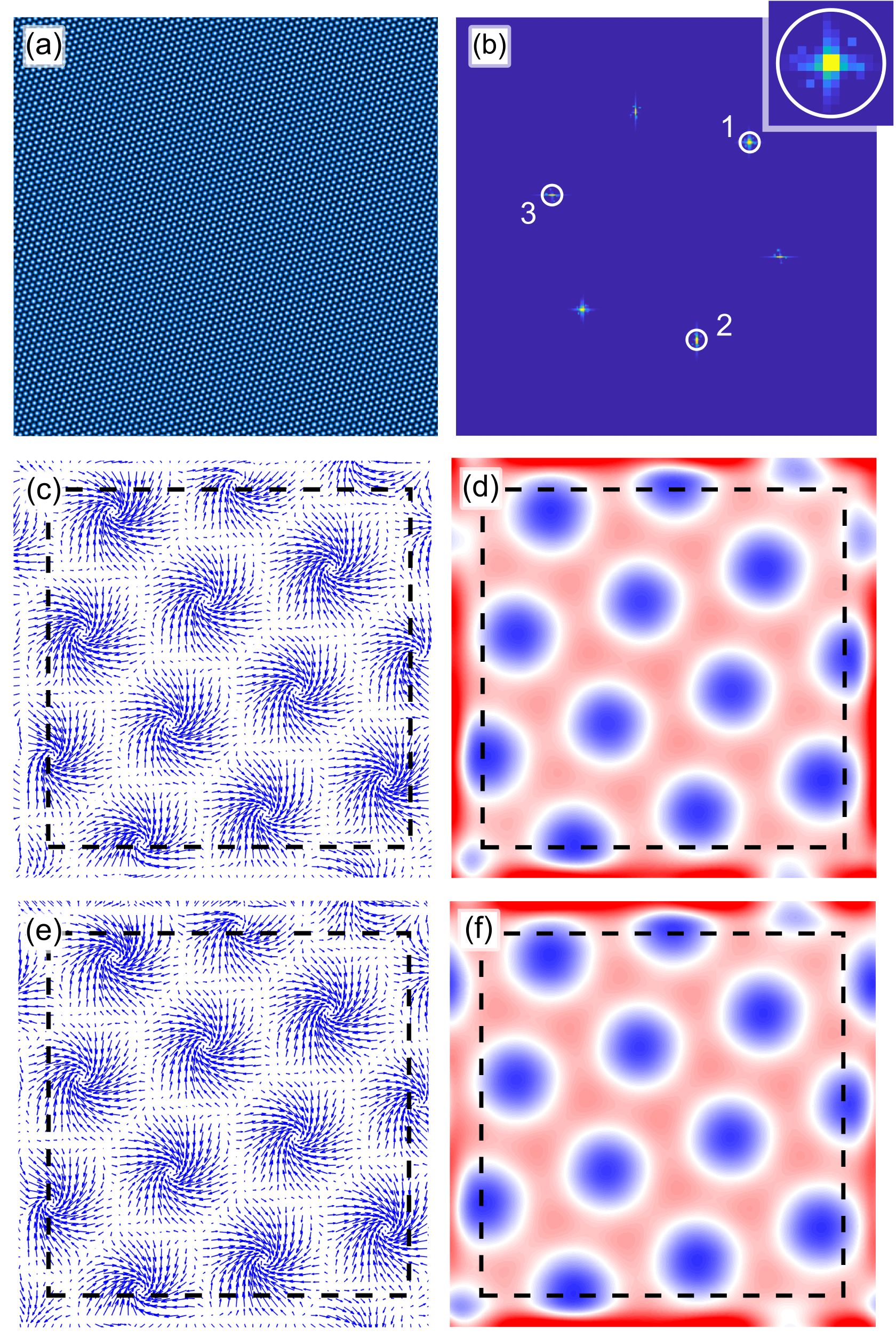}
    \caption{Chiral displacement field and periodic strain extracted from a calculated image which includes a real displacement field. (a) Topography of the distorted lattice. The absence of any visible domain structure is explained in the text. (b) FT of (a) showing the lattice peaks and their satellites arising due to the periodic distortion. The numbers identify three lattice peaks available for the GPA. The inset shows a zoom into one of the lattice peaks with its satellites. (c) Chiral displacement field and (d) biaxial strain field extracted by GPA using all three lattice peaks in the FT. (e) Chiral displacement field and (f) biaxial strain field extracted by GPA using only the lattice peaks \#1 and \#2. Note that the results do not depend on the choice of lattice peaks combination (see also \SF[9]).}
    \label{fig:periodic_displacement}
\end{figure}

Let us finally consider the case of an STM image with a real spatial distortion. To this end, we use a bespoke periodic chiral displacement field to distort a perfect lattice as described in \suppsec[]~\Romannum{6}. This chirality is hardly seen in the corresponding topography (\F[\ref{fig:periodic_displacement}](a)), which is not surprising, as a pure distortion acts on the phase of the signal while the peak-to-peak amplitude remains intact. This remains true even when we apply a distortion field so large that the atomic displacements become visually perceptible. However, the periodic chiral distortion is unambiguously seen in the corresponding FT (\F[\ref{fig:periodic_displacement}](b)) as satellite peaks surrounding the lattice peaks. Their positions correspond to the $\v{q}_{Li}\pm\v{q}_{Mj}$ wavevectors, where $\v{q}_{Li}$ and $\v{q}_{Mj}$ are the wavevectors of the undistorted lattice and of the periodic displacement field, respectively. The magnified lattice peak region in the inset of \F[\ref{fig:periodic_displacement}](b) reveals asymmetric satellite peaks, as expected in the presence of a periodic displacement field.

Applying GPA to this distorted image, we recover precisely the displacement field we used to generate the distorted image (\F[\ref{fig:periodic_displacement}](c) and we can extract the corresponding strain (\F[\ref{fig:periodic_displacement}](d)). This result demonstrates the suitability of GPA to access genuine strain fields. Since we are dealing here with real strain and not an artifact, the result should not depend on whether we solve the over-constrained problem using all three lattice peaks as in \F[\ref{fig:periodic_displacement}](c) and (d), or whether we solve it using any pair of lattice peaks. Indeed, the results in \F[\ref{fig:periodic_displacement}](e) and (f), obtained using only two lattice peaks, are exactly the same, and this is true for any pair of lattice peaks as shown in \SF[9].

The last example suggests a possible method to disentangle real strain from artifacts by comparing the GPA results obtained with two different pairs of lattice peaks. Unfortunately, this may not work in experimental data of moir\'e systems, as the satellite peaks can have a combined origin. They may result from a combination of moir\'e wavevector mixing and periodic distortion. Without a suitable strategy --which we are not yet aware of-- to disentangle these contributions, GPA-based strain analysis is ambiguous in moir\'e systems: one may find a periodic strain where there is absolutely no strain present in the system. The only reliable statement possible is in the case of pure deformation, where GPA analysis gives exactly the same result irrespective of the choice of lattice peaks for the analysis. Whenever different results are found, it is impossible to make any strong claim about the presence or absence of strain. This problem is formally very similar to the one discussed in the context of scanning transmission electron microscopy of composite materials, where a variation in the unit cell composition leads to an apparent strain at the interface between two compounds although there is no strain \cite{Peters2015}. 

A further upshot of our analysis is to highlight a risk often overlooked when applying GPA-based distortion corrections (notably the popular Lawler-Fujita method \cite{Lawler2010}) to scanning probe images. GPA has to be deployed with care and is only applicable when there is no \textit{contamination} in the vicinity of the lattice peaks, which otherwise leads to deceiving apparent distortions; correcting for them is likely to introduce spurious features that are not genuinely representative of the sample under investigation.

\section{Conclusion}

We have analyzed possible pitfalls of geometric phase analysis to extract periodic deformation and displacement fields from scanning tunneling microscopy images when several atomic-scale lattices contribute to the imaging contrast. While our focus is mainly on the moir\'e patterns observed on heterostructures, the conclusions are general. We demonstrate the possible misinterpretation of high resolution images whose contrast is determined by mixed signals from at least two lattices, one of which may also be a charge density wave. They can mimic deformation, chirality and other displacement and strain fields, despite the system under investigation being perfectly undistorted and relaxed. 

\section{Methods}

The samples and data presented in \F[\ref{fig:FT_heterostructures}] were prepared and measured at the University of Geneva as described in \cite{Sun2022, Sun2022_2}, \cite{Martinez2018},\cite{Pushkarna2023} for panel (a), (b) and (c) respectively. 

\section*{ACKNOWLEDGEMENTS}
We acknowledge A. Scarfato, L. Sun, and B. Horv\'ath for stimulating scientific discussions. This work was supported by the Swiss National Science Foundation (Division II Grant No. 182652). 

\bibliographystyle{ieeetr}
\bibliography{bibliography}

\end{document}


\title{Supplementary Material for "Delusive chirality and periodic strain pattern in moir\'{e} systems"}

\author{\'{A}rp\'{a}d P\'{a}sztor}
\email{arpad.pasztor@unige.ch}
\affiliation{DQMP, Université de Genève, 24 Quai Ernest Ansermet, CH-1211 Geneva, Switzerland}
\affiliation{These authors contributed equally to this work.}

\author{Ishita Pushkarna}
\affiliation{DQMP, Université de Genève, 24 Quai Ernest Ansermet, CH-1211 Geneva, Switzerland}
\affiliation{These authors contributed equally to this work.}

\author{Christoph Renner}
\email{christoph.renner@unige.ch}
\affiliation{DQMP, Université de Genève, 24 Quai Ernest Ansermet, CH-1211 Geneva, Switzerland}

\maketitle

\tableofcontents

\newpage
\section{Aggregated signal of the sum of plane waves}
\label{sup_sec:sum_of_plane_waves}

In this section, we examine simple cases to enlighten the meaning of Equation~(3) and Equation~(4). First, we consider only two q-vectors: $\v{q}_{10}=\v{q}_L$ and $\v{q}_{11}=\v{q}_L+\v{q}_M$; and with amplitudes: $A_{10}(\v{r})=\alpha A_0$ and $A_{11}(\v{r})=A_0$. By substituting these into Equations~(3) and (4), we find:
\begin{equation}
            A(\v{r})=A_0\sqrt{\alpha^2+1+2\alpha\cos{(\v{q}_M\v{r})}}
\end{equation}
\begin{equation}
\varphi(\v{r})=\arctan\left(\frac{\sin (\v{q}_M\v{r})}{\alpha+\cos(\v{q}_M\v{r})} \right),
\end{equation}
which clearly shows that Equation~(3) and Equation~(4) describe an amplitude-modulated signal with spatially evolving phase, given by $\varphi(\v{r})$ with respect to a $T_0(\v{r})=\sin(\v{q}_L\v{r})$ reference signal. The high-amplitude domains of this \textit{spatial beating} signal appear when the constituent waves interfere constructively, i.e. when they are in phase: 
\begin{equation}
\begin{split}
               \v{q}_{10}\v{r}+n2\pi=\v{q}_{11}\v{r}\\
            n2\pi=\v{q}_{M}\v{r}, 
\end{split}
\end{equation}
which shows that the period of the amplitude and the phase modulations is given by $\v{q}_M$. 

\begin{figure}[htp]
    \centering
    \includegraphics[]{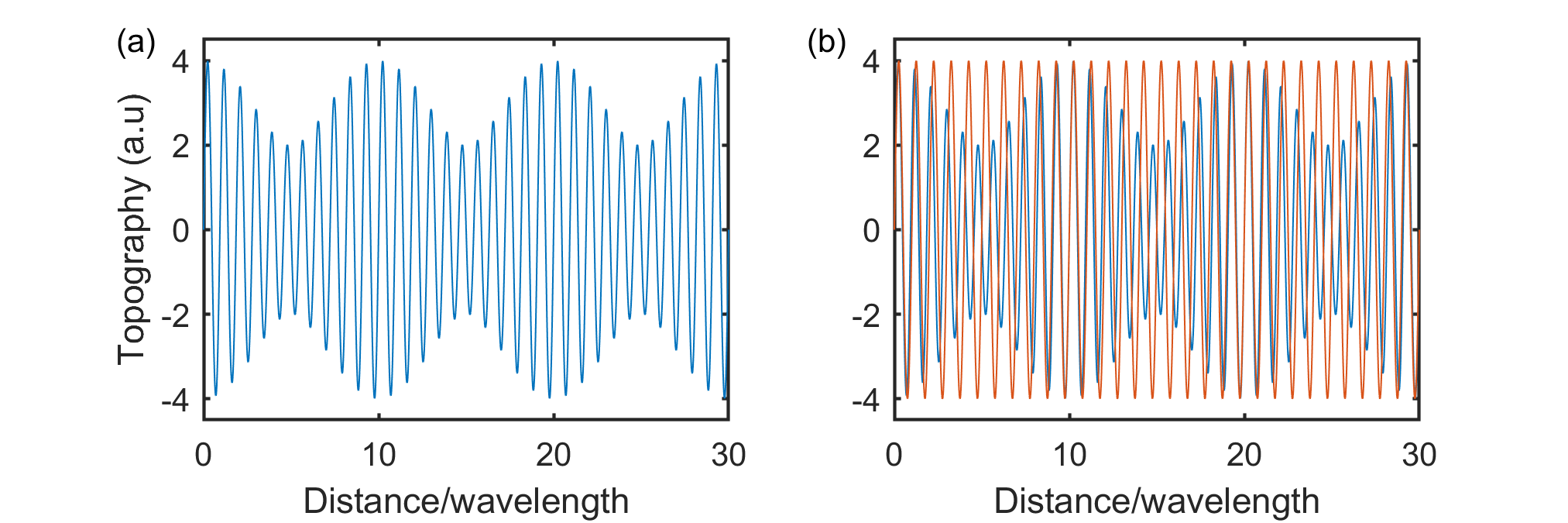}
    \caption{Sum of two sine waves in one dimension. (a) Sum of two sine functions showing the amplitude-modulated signal. (b) To highlight the phase relation we plot the sum of two sines (blue) and the $\sin(q_Lx)$ reference signal (red) with matching amplitude for easy comparison.}
    \label{fig:sup_beating1}
\end{figure}

 We plot the sum of two one-dimensional sine waves showing the amplitude-modulated beating signal in \SF[\ref{fig:sup_beating1}](a). The phase relation is highlighted in \SF[\ref{fig:sup_beating1}](b) where we plot the beating signal together with the $\sin({q}_Lx)$ reference signal whose amplitude is matched for a better comparison.

\begin{figure}[h]
    \centering
    \includegraphics[width=0.65\columnwidth]{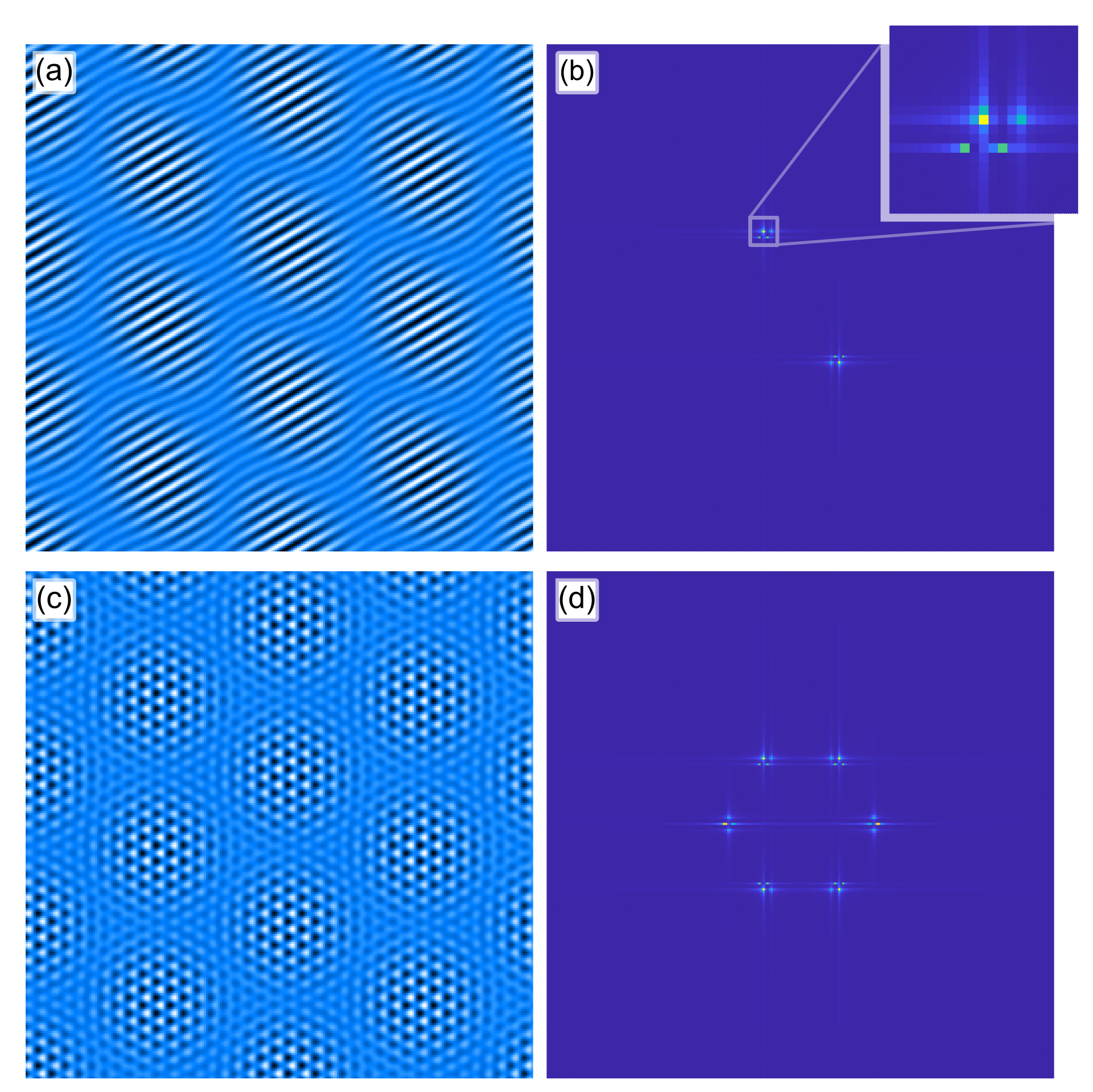}
    \caption{(a) Amplitude modulated topography, constructed by the sum of plane waves in 2D. (b) Fourier transform of (a) showing two lattice peaks, surrounded by satellite peaks. (c) Simulated topography with three lattice plane waves and their satellites, along with its FT in (d).}
    \label{fig:sup_beating_lattice_2d}
\end{figure}

\SF[\ref{fig:sup_beating_lattice_2d}] shows the phenomenon of spatial beating in two dimensions. In panel (a) we show the simulated topography that corresponds to the sum of plane waves with wavevectors giving a lattice peak along with three satellite peaks in the Fourier transform (FT) in panel (b) (see inset for a zoom into the FT). We clearly see in panel (a) that the amplitude of this plane wave is modulated along a long wavelength lattice defined by wavevectors pointing from the lattice to the satellite peaks. In panel (c) we show a simulated topography corresponding to plane waves with $2\pi/3$ rotated lattice peak in Fourier space (panel (d)) and their satellite peaks constructed the same way as in (a) and (b). 

\newpage
\clearpage

\section{Displacement and strain field}
\label{sup_sec:Displacement_field}
In this section, we discuss how to extract a displacement and strain field from topographic images. We consider a topography ($T(\v{r})$) where the imaging distortions and local strain lead to a change in the original positions: $\v{r}'=\v{r}+\v{u}(\v{r})$.

We assume that the topography is composed of the sum of harmonic modulations: $T(\v{r}')=\sum_iT_i(\v{r}')=\sum_i\sin(\v{q}_i\v{r}')$. Taking only two modulations we can write:
\begin{equation}
    T_1(\v{r}')=\sin({\v{q}_1\v{r}+\v{q}_1\v{u}(\v{r}))}=\sin({\v{q}_1\v{r}+\varphi_1(\v{r}}))
\end{equation}
\begin{equation}
    T_2(\v{r}')=\sin({\v{q}_2\v{r}+\v{q}_2\v{u}(\v{r}))}=\sin({\v{q}_2\v{r}+\varphi_2(\v{r}}))).
\end{equation}
with
\begin{equation}
\begin{split}
    \v{q}_1\v{u}(\v{r})=\varphi_1(\v{r}) \\
    \v{q}_2\v{u}(\v{r})=\varphi_2(\v{r}).
\end{split}
\end{equation}
Writing component by component we get,
\begin{equation}
\begin{split}
    q_{1x}u_x(\v{r})+q_{1y}u_y(\v{r})=\varphi_1(\v{r}) \\
    q_{2x}u_x(\v{r})+q_{2y}u_y(\v{r})=\varphi_2(\v{r}),
\end{split}
\end{equation}
which can be written in a matrix equation form as:
\begin{equation}
\v{Q}\v{U(\v{r})}=\vec{\varphi}(\v{r}),
\label{sup_eq:displacement_field1}
\end{equation}
where
\begin{equation}
\begin{split}
    \v{Q}=\begin{pmatrix}
    q_{1x} & q_{1y}\\
    q_{2x} & q_{2y}
    \end{pmatrix}\\
    \\       
    {\v{U}(\v{r})}=\begin{pmatrix}
    u_x(\v{r})\\
    u_y(\v{r}) 
    \end{pmatrix}\\
    \\
    \vec{\varphi}(\v{r})=\begin{pmatrix}
    \varphi_1(\v{r})\\
    \varphi_2(\v{r})
    \end{pmatrix}.\\
\end{split}
\end{equation}
The solution of this system of equations is:
\begin{equation}
\v{U(\v{r})}=\v{Q}^{-1}\vec{\varphi}(\v{r}).
\end{equation}

The equations above show that we need to consider (at least) two peaks in the FT in order to extract the displacement field, otherwise, we get an under-determined system of equations. On the other hand, we have not specified which two peaks. As long as they correspond to structural signals (atomic Bragg-peaks), we assume that the obtained displacement field represents structural strain and imaging artifacts.

After obtaining $\v{U(\v{r})}$ (either by the spatial lock-in or the real space fitting method \cite{Lawler2010,Benschop2021,Savitzky2017,Baggari2018,Pasztor2019}) we can correct the distortions of the topographic signal by mapping $T(\v{r}')$ to $T(\v{r}'-\v{u}(\v{r}))$ which can be done e.g by the \textit{imwarp} \cite{imwarp} function in Matlab. 

This approach is widely used in high-temperature superconductor research to correct imaging distortions accumulated over a long acquisition time. In this case, the atomic lattice has four (two)-fold symmetry with only two independent lattice peaks rendering the selection of peaks in the FT trivial for the analysis. On the other hand, in the transition metal dichalcogenides -- which are the main focus of this paper -- there are three independent lattice peaks and hence three independent choices for the set of lattice peaks that could be used for the analysis. 
 In principle, if the obtained phase field is due to real displacements any of the cases should lead to the same result, however in the main text and  \suppsec[]~\ref{sup_sec:2q} we show that when the obtained phase variation is not (purely) due to real distortion, using different peaks lead to very different results, e.g. artificial symmetry lowering, see \F[4] and \SF[\ref{fig:sup_2q_analytical}]{}. 

A modified method has been proposed (though, neither implemented nor compared to the original) by Benschop, Jong, Stepanov \textit{et al.} \cite{Benschop2021}, which involves using the phase of all three independent lattice peaks and solving the obtained over-determined system of equations in a least-square minimization sense. Mathematically, we start with:
\begin{equation}
\begin{split}
    \v{q}_1\v{u}(\v{r})=\varphi_1(\v{r}) \\
    \v{q}_2\v{u}(\v{r})=\varphi_2(\v{r})\\
    \v{q}_3\v{u}(\v{r})=\varphi_3(\v{r}).
\end{split}
\end{equation}
Writing component by component we get,
\begin{equation}
\begin{split}
    q_{1x}u_x(\v{r})+q_{1y}u_y(\v{r})=\varphi_1(\v{r}) \\
    q_{2x}u_x(\v{r})+q_{2y}u_y(\v{r})=\varphi_2(\v{r}) \\
    q_{2x}u_x(\v{r})+q_{2y}u_y(\v{r})=\varphi_2(\v{r})
\end{split}
\end{equation}
which can be written in a matrix equation form as:
\begin{equation}
\label{sup_eq:displacement_field2}
\v{Q}\v{U(\v{r})}=\vec{\varphi}(\v{r}),
\end{equation}
where
\begin{equation}
\begin{split}
    \v{Q}=\begin{pmatrix}
    q_{1x} & q_{1y}\\
    q_{2x} & q_{2y} \\
    q_{3x} & q_{3y}
    \end{pmatrix}\\
    \\       
    {\v{U}(\v{r})}=\begin{pmatrix}
    u_x(\v{r})\\
    u_y(\v{r}) 
    \end{pmatrix}\\
    \\
    \vec{\varphi}(\v{r})=\begin{pmatrix}
    \varphi_1(\v{r})\\
    \varphi_2(\v{r}) \\
    \varphi_3(\v{r})
    \end{pmatrix}\\.
\end{split}
\end{equation}

Supplementary equation (\ref{sup_eq:displacement_field2}) is formally the same as the previous supplementary equation (\ref{sup_eq:displacement_field1}) for the displacement field. However, this time $\vec{Q}$ is not invertible and we cannot find the solution as $\v{U(\v{r})}=\v{Q}^{-1}\vec{\varphi}(\v{r})$. Instead, at each pixel, we can search for the $\v{U(\v{r})}$ which minimizes the Euclidean norm of the $\v{Q}\v{U(\v{r})}-\vec{\varphi}(\v{r})$ vector, i.e. a solution in the minimum of least-squares sense. This is easily implemented in Matlab using the \textit{mldivide} (or backslash)\footnote{https://ch.mathworks.com/help/matlab/math/systems-of-linear-equations.html 28.08.2023} operation pixel by pixel.

The displacement vector field \v{U}(\v{r}) can be used to find the local distortion of the lattice and to define the strain tensor at each position whose elements are defined by the derivatives of the displacement field, as follows\cite{Landau1986,hytch1998}:
\begin{equation}
    \begin{split}
        \v{S}(\v{r})=\begin{pmatrix}
         s_{xx}(\v{r}) & s_{xy}(\v{r})\\
         s_{yx}(\v{r}) & s_{yy}(\v{r})
        \end{pmatrix}=
        \begin{pmatrix}
         \frac{\partial U_x(\v{r})}{\partial x} & \frac{1}{2}\left(\frac{\partial U_x(\v{r})}{\partial y}+\frac{\partial U_y(\v{r})}{\partial x}\right)\\
         \frac{1}{2}\left(\frac{\partial U_x(\v{r})}{\partial y}+\frac{\partial U_y(\v{r})}{\partial x}\right) & \frac{\partial U_y(\v{r})}{\partial y}
        \end{pmatrix},
    \end{split}
\end{equation}
where $U_x(\v{r)}$ and $U_x(\v{r)}$ are the $x$ and $y$ component of the displacement field at position $\v{r}$. The normal strains in $x$ and $y$ directions interfere with each other and lead to an isotropic strain which is defined by $\epsilon_{bi}=(s_{xx}+s_{yy})/2$, commonly referred to as biaxial strain in the literature \cite{Walkup2018,Gao2018}, whereas the anisotropic (uniaxial) strain is defined as $\epsilon_{uni}=(s_{xx}-s_{yy})/2$. 

\newpage
\section{Geometric phase analysis using the spatial lock-in method}
\label{sup_sec:spatial_lockin}
To extract the geometric phase ($\varphi_1(\v{r})$) of the
$T(\v{r})=A(\v{r})\sin({\v{q}_1\v{r}+\varphi_1(\v{r}}))$ signal we first define a lock-in signal:
\begin{equation}
\begin{split}
    s_1(\v{r})&=\cos({\v{q}_1\v{r}}) \\
    s_2(\v{r})&=\sin({\v{q}_1\v{r}}),
\end{split}
\end{equation}
and then multiply the original signal with $s_1$ and $s_2$ at every location and low-pass filter:
\begin{equation}
\begin{split}
\Tilde{T}_1(\v{r})=T(\v{r})s_1(\v{r})=A(\v{r})\sin({\v{q}_1\v{r}+\varphi_1(\v{r}}))\cos({\v{q}_1\v{r})}=\\
=A(\v{r})\frac{1}{2}(\sin(\v{q}_1\v{r}+\varphi_1(\v{r})+\v{q}_1\v{r})+\sin(\v{q}_1\v{r}-\v{q}_1\v{r}+\varphi_1(\v{r})))=\\ \textit{after low-pass filtering}=A(\v{r})\frac{1}{2}\sin(\varphi_1(\v{r})).
\end{split}
\end{equation}
Similarly for $s_2$:
\begin{equation}
\begin{split}
\Tilde{T}_2(\v{r})=T(\v{r})s_2(\v{r})=A(\v{r})\sin({\v{q}_1\v{r}+\varphi_1(\v{r}}))\sin({\v{q}_1\v{r})}=\\
=A(\v{r})\frac{1}{2}(\cos(\v{q}_1\v{r}+\varphi_1(\v{r})-\v{q}_1\v{r})-\cos(\v{q}_1\v{r}+\varphi_1(\v{r})+\v{q}_1\v{r}))=\\
\textit{after low-pass filtering}=A(\v{r})\frac{1}{2}\cos(\varphi_1(\v{r}).)
\end{split}
\end{equation}
Then by taking the ratio
\begin{equation}
\begin{split}
\Tilde{T}_1(\v{r})/\Tilde{T}_2(\v{r})=\tan(\varphi_1(\v{r}))
\end{split}
\end{equation}
we can obtain the geometric phase from the (four-quadrant) arctangent:
\begin{equation}
\begin{split}
\arctan(\Tilde{T}_1(\v{r})/\Tilde{T}_2(\v{r}))=\varphi_1(\v{r}).
\end{split}
\end{equation}

\newpage

\section{Various cases within the sum of plane waves model}

\subsection{Numerical evaluation of the analytical form using three phases to determine the displacement field}

\begin{figure}[htp]
    \centering
    \includegraphics[width=0.75\columnwidth]{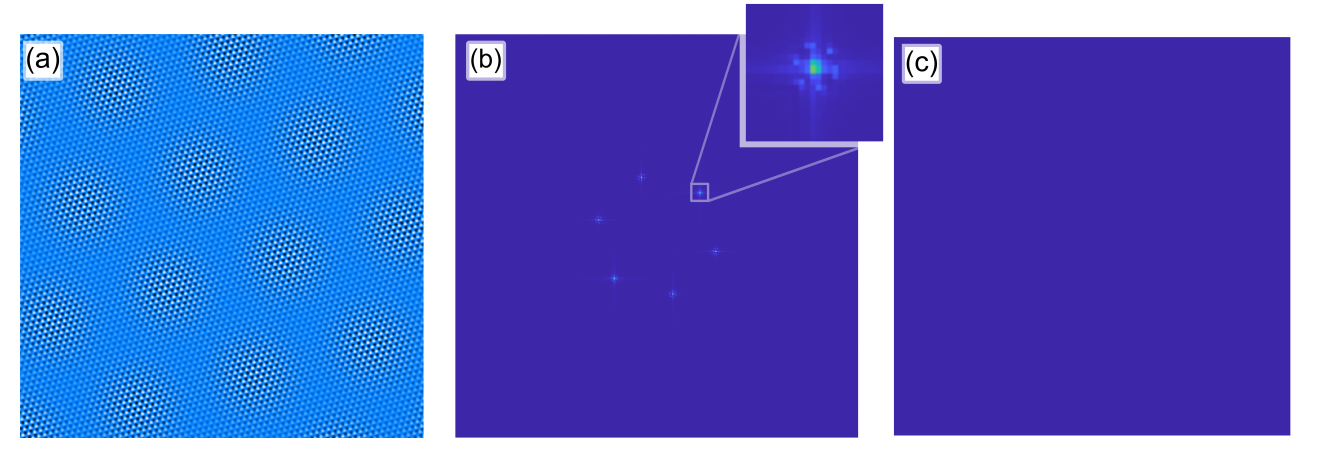}
    \caption{We find an amplitude-modulated topography (a) in the case when all the satellite peaks have the same intensity (b) (FT of (a)), but the displacement field (c) is identically zero.}
    \label{fig:sup_all_peaks_same}
\end{figure}

\begin{figure}[htp]
    \centering
    \includegraphics[width=0.75\columnwidth]{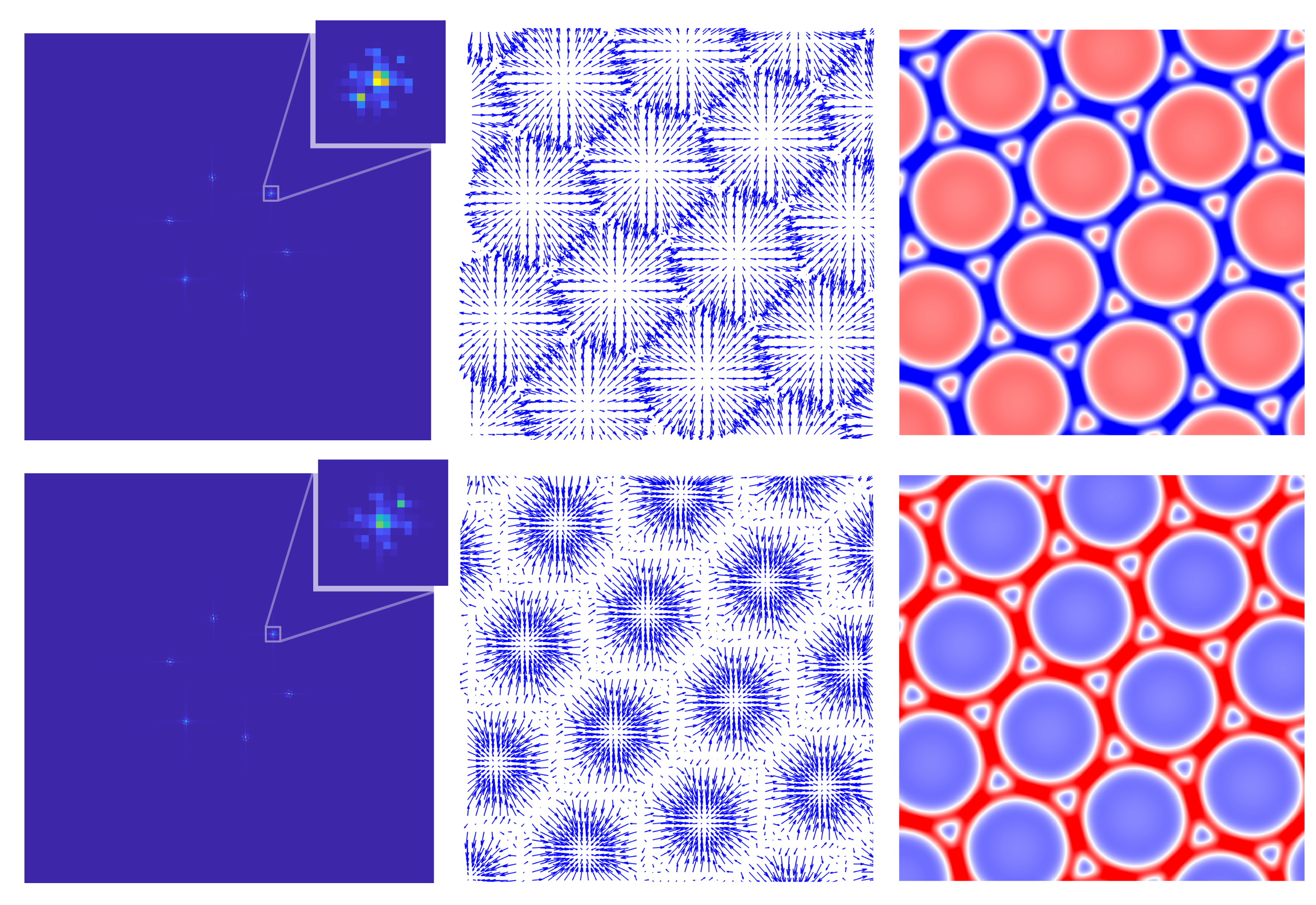}
    \caption{If one satellite peak is stronger than the others and is parallel with the lattice peak there is a radial displacement field: pointing away from (towards)  the middle of the domains. In each row, from left to right: FT of the real-space topography (inset: zoom into the FT around one of the lattice peaks), displacement vector field, and biaxial strain map. }
    \label{fig:sup_one_stronger_parallel}
\end{figure}

\begin{figure}[htp]
    \centering
    \includegraphics[width=0.75\columnwidth]{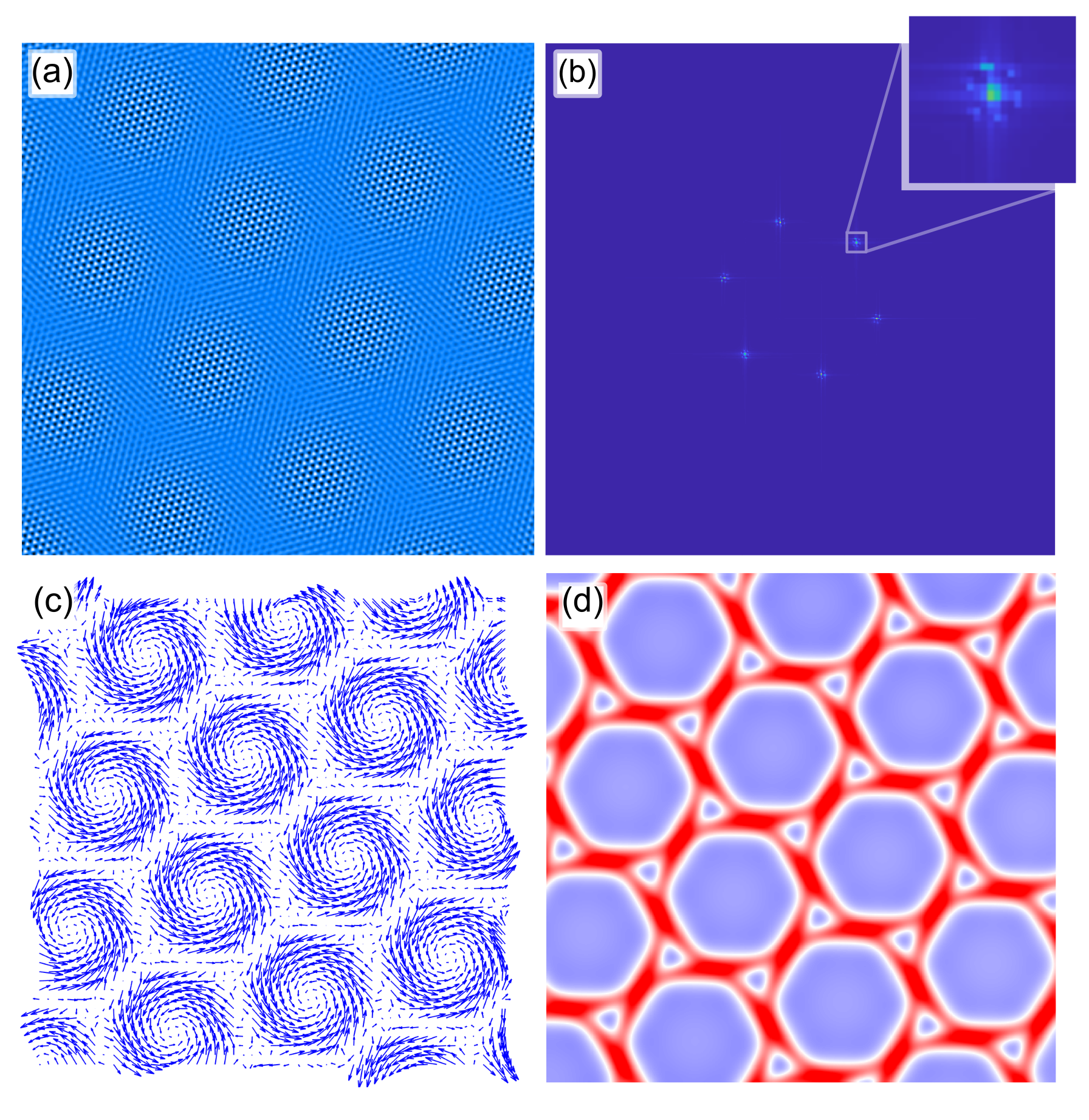}
    \caption{If one satellite peak is stronger than the others and is not parallel with the lattice peaks there is a chiral displacement and a periodic strain field. The direction of the local displacement field is opposite if this peak corresponds to $\v{q}_{Li}+\v{q}_{Mj}$ or $\v{q}_{Li}-\v{q}_{Mj}$. (a) Amplitude-modulated topography. (b) FT of (a) with an inset zooming on the vicinity of one of the lattice peaks to show that this case corresponds to $\v{q}_{Li}-\v{q}_{Mj}$ being stronger compared to \F[2]. (c) the vorticity of the chiral displacement field is changing compared to \F[2]. (d) Biaxial strain map, with inverted contrast compared to \F[2]. }
    \label{fig:sup_opposite_stronger}
\end{figure}

\newpage
\clearpage

\subsection{Application of the spatial lock-in method on images generated in the sum of plane wave model using three phases to determine the displacement field}

In all cases we find the same as with the evaluation of the analytical formula, apart from the borders where the Gaussian kernel is not yet entirely within the image. In the displacement vector field and biaxial strain maps, the dashed black squares delimit the region $3\sigma$ (in real space) away from the edges of the images, where the filtering kernel is entirely within the image and edge effects disappear.

\begin{figure}[htp]
    \centering
    \includegraphics[width=\columnwidth]{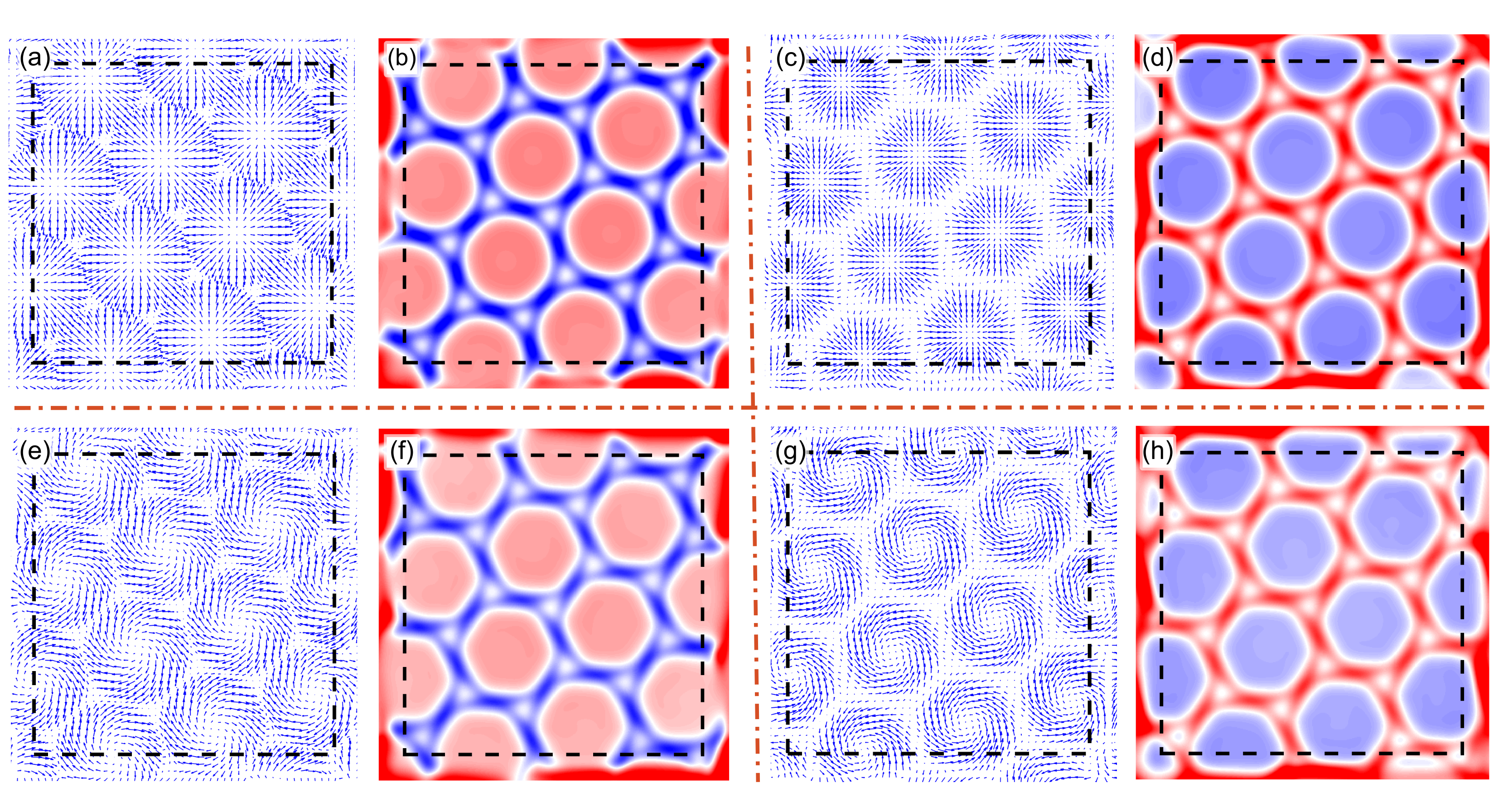}
    \caption{Displacement and biaxial strain fields extracted using the spatial-locking method in the same cases that we studied analytically. (a)-(d) One peak is stronger and parallel with the lattice peak, radial strain (\SF[\ref{fig:sup_one_stronger_parallel}]). (e) and (f) One peak is stronger (\F[2]). (g) and (h) The opposite peak is stronger (\SF[\ref{fig:sup_opposite_stronger}]). }
    \label{fig:sup_SL_all}
\end{figure}

\newpage
\clearpage
\section{Analytical result for the case discussed in Figure~3}

\begin{figure}[htp]
    \centering
    \includegraphics[width=0.4\columnwidth]{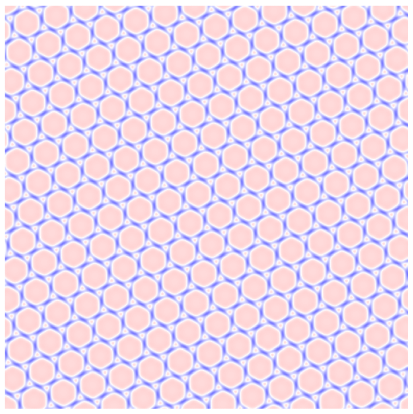}
    \caption{Strain field calculated using the analytical form of the phase for the case discussed in Figure~3 of the main text.  }
    \label{fig:sup_analytical_result_for_increased_moire}
\end{figure}

\section{Generation of chiral displacement fields}
\label{sup_sec:generate_chiral_dfield}
In order to test the effect of a periodic chiral displacement field on a perfect single lattice, we first generate a chiral displacement field and then distort the lattice accordingly. In practice, we first create a base lattice as the sum of three plane waves with q-vectors rotated $2\pi/3$ with respect to each other. For the displacement field, we start again from the sum of three plane waves, however this time with q-vectors which are shorter than for the base lattice; their length and direction can be selected on demand. Then we take the gradient of this field to obtain a vector field. Finally, we locally rotate the vectors of this field around the out-of-plane axis by a $\gamma$ angle to obtain a chiral displacement field. With this method, we can continuously tune the displacement field from fully radial (pointing toward/away from the domain center) through chiral (whirling around the domains) to entirely tangential (perpendicular to the direction toward the domain center) by choosing the appropriate $\gamma$. Finally, we can apply this displacement field (e.g using the \textit{imwarp} \cite{imwarp} function in Matlab) to the base lattice to generate the distorted lattice. 

\newpage
\clearpage
\section{Using different pairs of lattice peaks to extract the displacement field}
\label{sup_sec:2q}

\subsection{In case of phase fields obtained analytically}
\begin{figure}[htp]
    \centering
    \includegraphics[width=0.9\columnwidth]{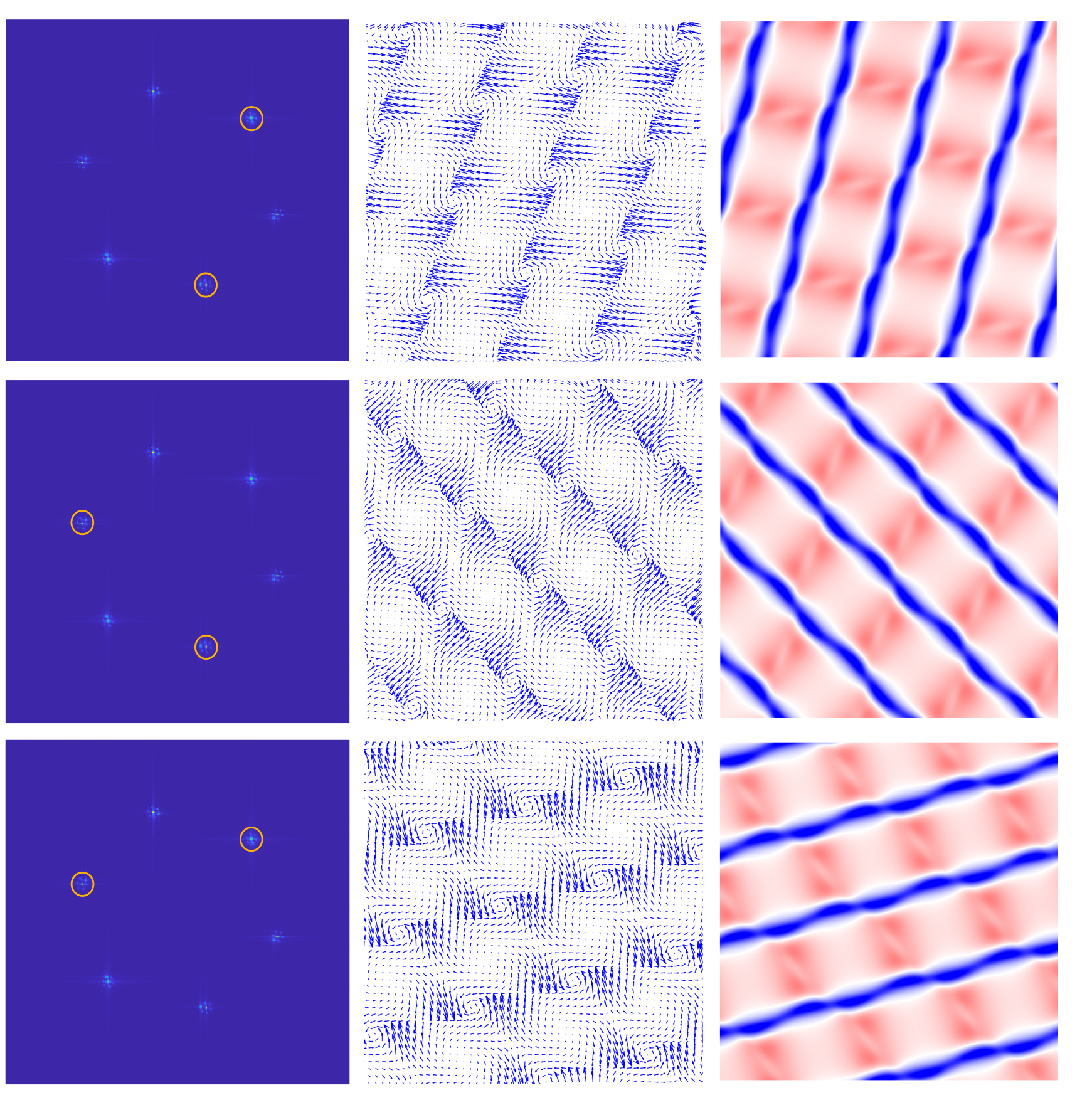}
    \caption{Even if the phase fields are obtained by evaluating the analytical formula, the obtained displacement and strain fields depend on the pair of lattice peaks chosen for the calculation.  In each row, left to right: FT of the real-space topography with orange circles marking the peaks whose phase was used to calculate the displacement field, the displacement vector field, and the biaxial strain field that we obtained in the analysis.}
    \label{fig:sup_2q_analytical}
\end{figure}

\newpage
\clearpage

\subsection{In case of a real chiral displacement field}
\begin{figure}[htp]
    \centering
    \includegraphics[width=0.77\columnwidth]{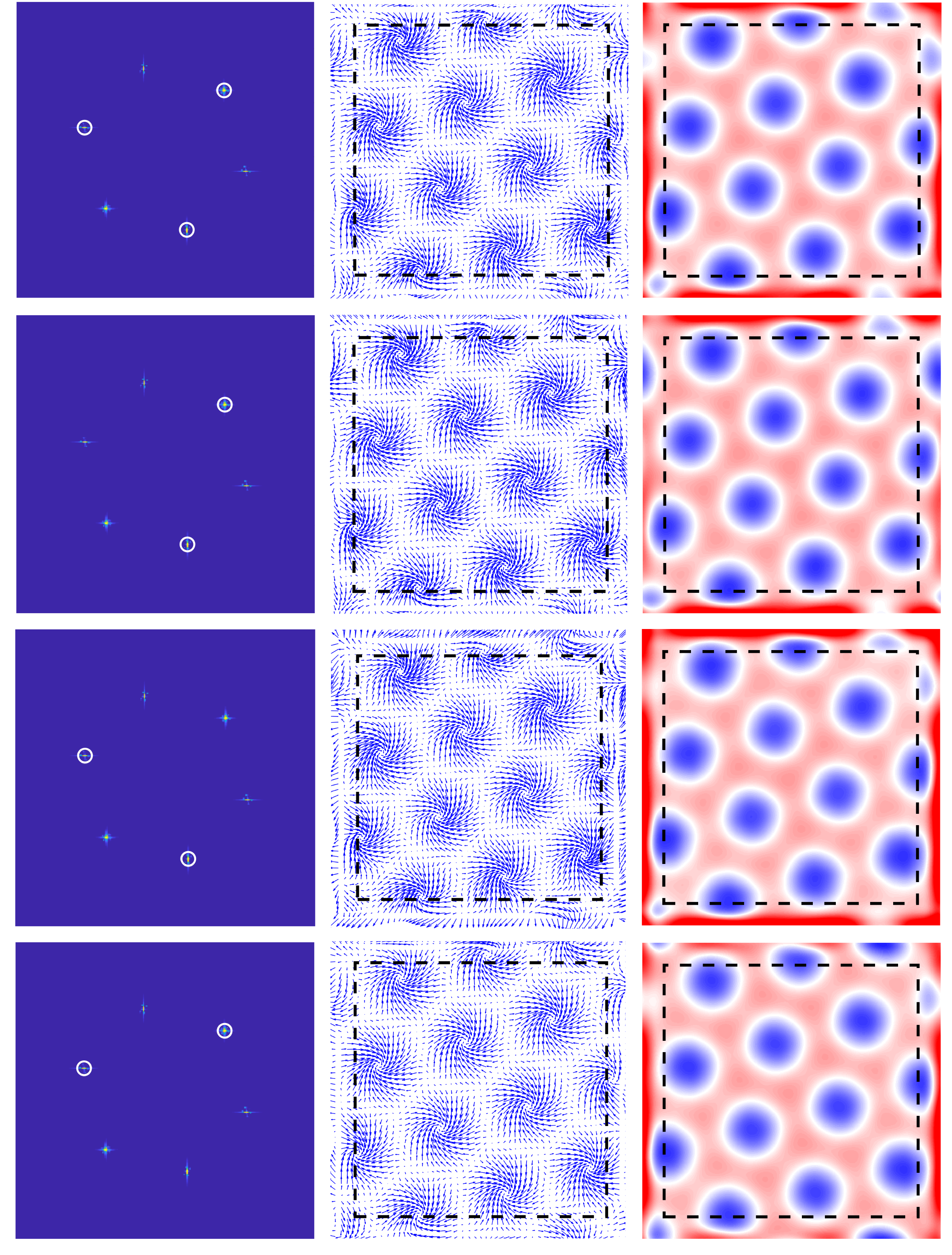}
    \caption{When the satellite peaks are due to real periodic displacements, the displacement and strain fields do not depend on which pair of lattice peaks we use for the analysis.  In each row, left to right: FT of the real-space topography with white circles marking the peaks that we use for the GPA, the displacement vector field, and the biaxial strain field that is obtained in the analysis.}
    \label{fig:sup_2q_chiral_displacement}
\end{figure}

\newpage
\clearpage
\section{Sum of sine functions}
\label{sup_sec:maths}

A sum of sine functions can be always written as a single sine function \footnote{Weisstein, E. W. (n.d.). Harmonic Addition Theorem. Mathworld.Wolfram.com. Retrieved August 28, 2023, from https://mathworld.wolfram.com/HarmonicAdditionTheorem.html} as:
\begin{equation}
    \sum_i a_i \sin(x + \varphi_i) = a \sin(x + \varphi)
\end{equation}
where
\begin{align}
     a^2 &= \sum_{i,j}a_i a_j \cos(\varphi_i - \varphi_j)\\ 
    \tan\varphi&= \frac{\sum_i a_i \sin\varphi_i}{\sum_i a_i \cos\varphi_i}.
\end{align}

\bibliographystyle{ieeetr}
\bibliography{bibliography}